Transport properties and microstructure of mono- and seven-core wires of FeSe$_{1-x}$Te$_x$ superconductor by Fe-diffusion powder-in-tube method

Toshinori Ozaki[1,3], Keita Deguchi[1,2,3], Yoshikazu Mizuguchi[1,2,3], Yasuna Kawasaki[1,2,3], Takayoshi Tanaka[1], Takahide Yamaguchi[1,3], Shunsuke Tsuda[1,3], Hiroaki Kumakura[1,2,3], Yoshihiko Takano[1,2,3]

[1] National Institute for Materials Science, 1-2-1 Sengen, Tsukuba, Ibaraki 305-0047, Japan

[2] University of Tsukuba, 1-1-1Tennnodai, Tsukuba, Ibaraki 305-0047, Japan

[3] JST, Transformative Research-project on Iron Pnictides, 1-2-1 Sengen, Tsukuba, Ibaraki 305-0047, Japan

(Abstract)

We report the successful fabrication of mono- and seven-core superconducting wires of FeSe$_{1-x}$Te$_x$ using an *in-situ* Fe-diffusion process based on the powder-in-tube (Fe-diffusion PIT) method. The reacted layer in these wires were found to have composite structure with composition nearly FeSe and FeTe for the inner and outer layers, although a single layer of composition FeSe$_{0.5}$Te$_{0.5}$ was supposed to be formed. The self-field transport $J_c$ values at 4.2 K were found to be 226.2 A/cm$^2$ and 100.3 A/cm$^2$ respectively for mono- and seven-core wires. The $J_c$'s of mono- and seven-core wires dropped rapidly at low fields and then showed a gradual decrease with increasing magnetic fields. In addition, the seven-core wire showed higher $J_c$ than the mono-core wire under higher magnetic fields, indicating that the seven-core wire of FeSe$_{1-x}$Te$_x$ superconductor using Fe-diffusion PIT method is advantageous for the superconducting-wire application under high magnetic fields.

I.   **INTRODUCTION**

The discovery of superconductivity in iron oxypnictides[1] has generated strong interest in understanding the fundamental properties as well as pursuing potential applications. To date, several families of iron-based superconductors have been discovered[2-5] and transition temperature $T_c$ has been raised above 50 K[1,6]. Among these iron-based superconductors, FeSe with $T_c$ = 8 K is an important material in elucidating the mechanism of superconductivity in iron-based superconductor owing to it's simple crystal structure composed solely of stacked iron-chalcogenide layers along the $c$-axis[4]. The $T_c$ of FeSe is found to increase up to 15 K by the partial substitution of Te or S for Se[7-10]. The $T_c$ also shows a dramatic increase with applied external pressure reaching values as high as 37 K[11-13]. Earlier studies have shown that the upper critical field ($H_{c2}$) of FeSe$_{0.5}$Te$_{0.5}$ is as high as ~50 T[14]. The starting materials for these chalcogenides are less-toxic compared to the FeAs-based compounds. These aspects make the iron chalcogenides potential candidates for applications among the new iron-based superconductors.

We have succeeded in observing the transport critical current density $J_c$ of FeSe$_{1-x}$Te$_x$ tape[15] using an *in-situ* Fe-diffusion process based on the powder-in-tube (Fe-diffusion PIT) method. The interesting aspect of this process is that the Fe sheath plays the role of not only the sheath but also the raw materials for synthesizing the superconducting phase. This process is very simple and advantageous for fabricating superconducting wires. In our previous report on the superconducting tapes of the FeSe$_{1-x}$Te$_x$[15], we have adopted an annealing temperature 400°C. Such a low temperature was used to prevent evaporation of chalcogen during a treat treatment. However, in order to enhance $J_c$, much better grain connectivity is required, which calls for higher sintering temperatures. Therefore, we decided to sinter the wires at 700°C, which is the temperature used for the synthesis of polycrystalline bulk sample of FeSe$_{1-x}$Te$_x$. For high temperature sintering, the effective diffusion-distance of Fe needs to be shortened so that Fe reacts with chalcogen before it evaporates. For this purpose, the wires were thinned down with a wire-drawing die. We also fabricated the multi-core wires, where the cross-sectional area of the individual wires are further smaller leading to better possibility for the reaction without the escape of the chalcogen atoms. This paper provides the details on the fabrication and the superconducting properties of mono- and seven-core wires of FeSe$_{1-x}$Te$_x$ superconductor using the Fe-diffusion PIT method.

## II.  EXPERIMENTAL

We prepared the SeTe precursor with a ratio of 1 : 1 by solid-state reaction. The obtained precursor was ground and filled into a pure Fe tube with an outer diameter of 6.2 mm, an inner diameter of 3.5 mm and a length of 48 mm. The tube was rolled into a rectangular rod of about 2.5 mm in size using groove rolling. After that, in order to reduce its diameter, it was drawn into a wire of 1.1 mm in diameter using wire-drawing die. This wire was cut into pieces. Some wire pieces were used as sample of mono-core wires. A seven-core wire was produced by packing the unsintered seven pieces of the mono-core wires into another Fe tube. The seven-core composite was drawn into a wire with a final diameter of 2.0 mm. The seven-core wire was cut into short pieces. These mono- and seven-core wires were sealed inside a quartz tube evacuated and back-filled with argon gas. These sealed wires were heat-treated at 700 °C for 2 hours.

The microstructure of these wires was investigated with scanning electron microscope (SEM) and x-ray diffraction (XRD). An actual composition of the reacted layer was investigated using energy dispersive x-ray spectrometry (EDX). Transport critical currents ($I_c$) were measured for 4 cm-long wires by a standard four-probe resistive method at 4.2 K in magnetic fields. The magnetic field was applied perpendicularly to the wire axis. The criterion of $I_c$ definition was 1 $\mu$V/cm. The $J_c$ was obtained by dividing $I_c$ by the cross sectional area of the FeSe$_{1-x}$Te$_x$ core excluding the hole, which was measured by optical microscope.

## III. RESULTS AND DISCUSSION

Figure 1(a) and 1(b) show optical micrographs of the polished transverse cross section of the as-drawn mono- and seven-core wires, respectively. The cross sections of these wires show uniform deformation of the composite. Figure 2(a) and 2(b) show the polished transverse cross section of mono- and seven-core wires after heat treatment at 700°C for 2 hours. A reacted layer was observed on the inside wall of the Fe sheath and a hole was formed at the center of each core, where the TeSe precursor was filled before heat treatment. Theoretically, the volume contraction caused by the formation of FeSe$_{0.5}$Te$_{0.5}$ would not occur. Hence, the formation of the hole would be due to the evaporation of Se and Te.

Figure 3 shows the magnetic field dependence of transport $J_c$ of the mono- and seven-core FeSe$_{1-x}$Te$_x$ wires at 4.2 K. We succeeded in observing the transport $J_c$ for mono- and seven-core FeSe$_{1-x}$Te$_x$ wires. The self-field $J_c$ values for mono- and seven-core wire are as high as 226.2 A/cm$^2$ and 100.3 A/cm$^2$ at 4.2 K, respectively. For comparison, the $J_c$-$B$ curve at 4.2 K for the FeSe$_{1-x}$Te$_x$ tape data[15] reported previously is also plotted in Fig 3. From this comparison, it is seen that the $J_c$ value of mono-core wire is about 20 times higher than that of the FeSe$_{1-x}$Te$_x$ tape. $J_c$'s of mono- and seven-core wires showed a rapid decrease at low fields and then gradually decreased with increasing magnetic field. Furthermore, the seven-core wire exhibits higher $J_c$ than the mono-core wire in high magnetic fields, indicating that the seven-core wire prepared by the Fe-diffusion PIT method have a clear advantages for technological applications in high magnetic fields. However, the $J_c$ is about three orders of magnitude less than intra-grain $J_c$[16]. This implies the presence of weak links between grains. Higher $J_c$ values would be expected by enhancement of grain connectivity.

Temperature dependence of resistivity of the mono-core FeSe$_{1-x}$Te$_x$ wires under different applied magnetic fields is shown in Fig. 4. The resistivity at 0 T began to decrease at 11.6 K and drops to zero at 9.9 K. It is clear that the $\rho(T)$ curves are shifted to lower temperatures with increasing magnetic field without noticeably broadening. The transition width $\Delta T$ defined by the 90% and 10% points on $\rho(T)$ is less than 2 K. This behavior is similar to that of the low-temperature superconductors with small anisotropy[17,18]. We have estimated upper critical field ($\mu_0 H_{c2}$) and the irreversibility field ($\mu_0 H_{irr}$), using 90% and 10% of normal state resistivity, respectively. The $\mu_0 H_{c2}$ and $\mu_0 H_{irr}$ are plotted in the inset of Fig. 4 as a function of temperature. The $\mu_0 H_{irr}$ line is very close to the $\mu_0 H_{c2}$ line. These lines show an upturn curvature near 0 T. Such curve could be ascribed to the effect of excess Fe[19]. Above 2 T, both $\mu_0 H_{c2}(T)$ and $\mu_0 H_{irr}(T)$ lines show a linear curve with slopes of $d\mu_0 H_{c2}/dT = 2.6$ T/K and $d\mu_0 H_{irr}/dT = 2.5$ T/K. Linear extrapolation of the $\mu_0 H_{c2}(T)$ and $\mu_0 H_{irr}(T)$ data suggests $\mu_0 H_{c2}(0) \sim 27$ T and $\mu_0 H_{irr}(0) \sim 22$ T.

In order to identify the different phases formed by heat treatment, XRD analysis was performed for the reacted layer. Figure 5 shows the XRD pattern of the reacted layers obtained from the mono-core FeSe$_{1-x}$Te$_x$ wire. For comparison, calculated data of FeSe and FeTe are also shown in Fig. 5. It was found that the main peaks were identified as the FeSe and FeTe, and minor peak was identified as the hexagonal phase indicated with asterisk, although SeTe powder were mixed in a Se : Te = 1 : 1 atmoic ratio to form mixed composition FeSe$_{0.5}$Te$_{0.5}$. Table 1 shows lattice constants *a* and *c* of FeTe and FeSe formed in FeSe$_{1-x}$Te$_{x}$ wire calculated from the XRD pattern. The calculated lattice parameter of FeTe formed in the FeSe$_{1-x}$Te$_{x}$ wire was somewhat smaller compared to that of bulk[9,20]. In contrast, the calculated lattice parameter of FeSe formed in the FeSe$_{1-x}$Te$_{x}$ wire was somewhat higher compared to that of bulk[9,21]. The shrinkage of lattice parameter of FeTe might be due to the partial substitution of Se for the Te site. On the other hand, the extension of lattice parameter of FeSe is likely to arise from the partial substitution of Te at the Se site. In fact, the resistivity of the mono-core FeSe$_{0.5}$Te$_{0.5}$ wire dropped to zero at $T_c^{zero}$ = 9.9 K, indicating the partial substitution of Te for the Se site in FeSe because $T_c^{zero}$ observed for FeSe is only ~8 K[4,21].

Figure 6 displays a SEM image and the elemental mapping images for the polished longitudinal cross section of the mono-core FeSe$_{1-x}$Te$_{x}$ wire. The Fe distribution is homogeneous in reacted layer, which indicate that the Fe sheath reasonably supplied Fe to the reacted layer. It is found that the reacted layer had a composite structure of two main layers: The first layer, where Se is found to be distributed more densely near the Fe sheath and the second layer, where Te is distributed more densely close to the center of the Fe sheath. The weak concentration of Te was also observed at the interface between the Fe sheath and the reacted layer. Table 2 summarizes measured composition in points A-E in Fig. 6. EDX analysis of each spot showed that the reacted layer is composed of slightly Se-substituted FeTe layer and slightly Te-substituted FeSe layer. This is consistent with the result of XRD and elemental mapping images. The composition at the interface between the Fe sheath and superconducting core shows much less Fe and Te than $FeSe_{0.5}Te_{0.5}$. Given the result of XRD pattern, this might indicate that hexagonal $FeSe_{1-x}Te_x$ is formed at the interface. The slightly Se-substituted FeTe layer in the $FeSe_{1-x}Te_x$ wire might be an almost non-superconductor. We believe that the transport $J_c$ could be improved by increasing the FeSe layer, which might be accomplished by changing the atomic ratio of Te to Se packed into the Fe sheath. Presently work is in progress to check this hypothesis.

IV. CONCLUSION

We fabricated mono- and seven-core wires of FeSe$_{1-x}$Te$_x$ by the Fe-diffusion PIT method. The self-field transport $J_c$ values of 226.2 A/cm$^2$ and 100.3 A/cm$^2$ at 4.2 K were obtained for mono- and seven-core wire, respectively. The seven-core wire exhibits higher $J_c$ than the mono-core wire in high magnetic fields, indicating that the seven-core FeSe$_{1-x}$Te$_x$ wire could be promising for magnetic applications. The FeSe$_{1-x}$Te$_x$ wire showed practically no broadening of the resistive transition under magnetic fields. This behavior is similar to that of low-temperature superconductors with small anisotropy. The reacted layer was found to be composed of slightly Se-substituted FeTe layer and slightly Te-substituted FeSe layer, although Te powder and Se powder were mixed in a Te : Se = 1 : 1 atmoic ratio to form nominal composition FeSe$_{0.5}$Te$_{0.5}$. These results indicate that the optimization of the atomic ratio of Te to Se packed into Fe sheath and introduction of the pinning centers will help in realizing higher $J_c$ and $T_c$ values.

"Acknowledgments"

This work was supported in part by the Japan Society for the Promotion of Science (JSPS) through Grants-in-Aid for Scientific Research and Grants-in-Aid for JSPS Fellows and 'Funding program for World-Leading Innovative R&D on Science Technology (FIRST) Program'.

(references)

1) Y. Kamihara, T. Watanabe, M. Hirano, and H. Hosono, J. Am. Chem. Soc. **130**, 3296 (2008).

2) M. Rotter, M. Tegel, and D. Johrendt, Phys. Rev. Lett. **101**, 107006 (2008).

3) X. C. Wang, Q. Q. Liu, Y. X. Lv, W. B. Gao, L. X. Yang, R. C. Yu, F. Y. Li, and C. Q. Jin, Solid Stat Commun. **148**, 538 (2008).

4) F. C. Hsu, J. Y. Luo, K. W. Yeh, T. K. Chen, T. W. Huang, P. M. Wu, Y. C. Lee, Y. L. Huang, Y. Y. Chu, D. C. Yan, and M. K. Wu, Proc. Natl. Acad. Sci. U.S.A. **105,** 14262 (2008).

5) H. Ogino, S. Sato, K. Kishio, J. Shimoyama, T. Tohei, and Y. Ikuhara, Appl. Phys. Lett. **97**, 072506 (2010).

6) Z. A. Ren, W. Lu, J. Yang, W. Yi, X. L. Shen, Z. C. Li, G. C. Che, X. L. Dong, L. L. Sun, F. Zhou, and Z. X. Zhao, Chin. Phys. Lett. 25, 2215 (2008).

7) M. H. Fang, H. M. Pham, B. Qian, T. J. Liu, E. K. Vehstedt, Y. Liu, L. Spinu, and Z. Q. Mao, Phys. Rev. B **78**, 224503 (2008).

8) K. W. Yeh, T. W. Huang, Y. L. Huang, T. K. Chen, F. C. Hsu, P. M. Wu, Y. C. Lee, Y.

Y. Chu, C. L. Chen, J. Y. Luo, D. C. Yan, and M. K. Wu, Europhys. Lett. **84**, 37002 (2008).

9) Y. Mizuguchi, F. Tomioka, S. Tsuda, T. Yamaguchi, and Y. Takano, J. Phys. Soc. Jpn. **78**, 074712 (2009).

10) Y. Mizuguchi, F. Tomioka, S. Tsuda, T. Yamaguchi, and Y. Takano, Appl. Phys. Lett. **94**, 012503 (2009).

11) Y. Mizuguchi, F. Tomioka, S. Tsuda, T. Yamaguchi, and Y. Takano, Appl. Phys. Lett. **93**, 152505 (2008).

12) S. Medvedev, T. M. McQueen, I. A. Troyan, T. Palasyuk, M. I. Eremets, R. J. Cava, S. Naghavi, F. Casper, V. Ksenofontov, G. Wortmann, and C. Felser, Nat. Mater. **8**, 630 (2009).

13) S. Masaki, H. Kotegawa, Y. Hara, H. Tou, K. Murata, Y. Mizuguchi, and Y. Takano, J. Phys. Soc. Jpn. **78**, 063704 (2009).

14) S. Khim, J. W. Kim, E. S. Choi, Y. Bang, M. Nohora, H. Takagi, and K. H. Kim, Phys. Rev. B **81**, 184511 (2010).

15) Y. Mizuguchi, Keita Deguchi, S. Tsuda, T. Yamaguchi, H. Takeya, H. Kumakura,

and Y. Takano, Appl. Phys. Express **2**, 083004 (2009).

16) T. Taen, Y. Tsuchiya, Y. Nakajima, and T. Tamegai, Phys. Rev. B **80**, 092502 (2009).

17) A. Godeke, M. C. Fischer, A. A. Squitieri, P. J. Lee, and D. C. Larbalestier, J. Appl. Phys. **97**, 093909 (2005).

18) H. Kumakura, H. Kitaguchi, A. Matsumoto, H. Yamada, M. Hirakawa, and K. Tachikawa, Supercond. Sci. Technol. **18**, 147 (2005).

19) J. Ge, S. Gao, S. Shen, S. Yuan, B. Kang, and J. Zhang, Solid State Commun. **150**, 1641 (2010).

20) B. C. Sales, A. S. Sefat, M. A. McGuire, R. Y. Jin, D. Mandrus, and Y. Mozharivskyj, Phys. Rev. B **79**, 094521 (2009).

21) T. M. McQueen, Q. Huang, V. Ksenofontov, C. Felser, Q. Xu, H. Zandbergen, Y. S. Hor, J. Allred, A. J. Williams, D. Qu, J. Checkelsky, N. P. Ong, and R. J. Cava, Phys. Rev. B **79**, 014522 (2009).

(Captions)

Fig. 1 Optical microscope of transverse cross section for (a) mono-core wire and (b) seven-core wires before heat-treatment

Fig. 2 Optical microscope of transverse cross section for (a) mono-core wire and (b) seven-core wires after heat-treatment at 700°C and 2 hours

Fig. 3 Magnetic field dependence of transport $J_c$ at liquid helium temperature (4.2 K) for mono- and seven-core wires. $J_c$ was calculated by dividing the $I_c$ by the cross sectional area of reacted layer except for the holes. The magnetic field was applied perpendicular to the wire axis. The FeSe$_{1-x}$Te$_x$ tape was heat-treated at 400°C for 2 hours[15].

Fig. 4 Temperature dependence of resistivity for mono-core wires under magnetic fields up to 7 T. The inset shows temperature dependence of $H_{c2}$ and $H_{irr}$ determined from 90% and 10% points on the resistive transition curve.

Fig. 5 XRD diffraction pattern for reacted layer obtained from mono-core wire after heat treatment. Diffraction pattern of FeSe and FeTe is also shown in the lower panel of figure. Asterisk indicates hexagonal phase.

Fig. 6 Scanning electron microscopy (SEM) image and (b) Fe, (c) Se and (d) Te concentration mappings measured by EDX (energy-dispersive x-ray spectroscopy) on the polished longitudinal cross section for the mono-core wire.

Table 1 Lattice parameter of FeTe and FeSe formed in FeSe$_{1-x}$Te$_x$ mono-core wire.

|  | $a$-axis length [Å] | $c$-axis length [Å] |
| --- | --- | --- |
| FeTe | 3.8163(1) | 6.2308(6) |
| FeSe | 3.7800(7) | 5.5162(27) |

Table 2 Compositions detected by SEM-EDX for the points indicated in Fig. 6.

|   | Se | Te | Fe |
|---|---|---|---|
| A | 1.32 | 42.61 | 56.07 |
| B | 1.66 | 42.51 | 55.83 |
| C | 48.79 | 1.41 | 49.80 |
| D | 40.40 | 17.47 | 42.13 |
| E | 0 | 0 | 100 |

(atm%)

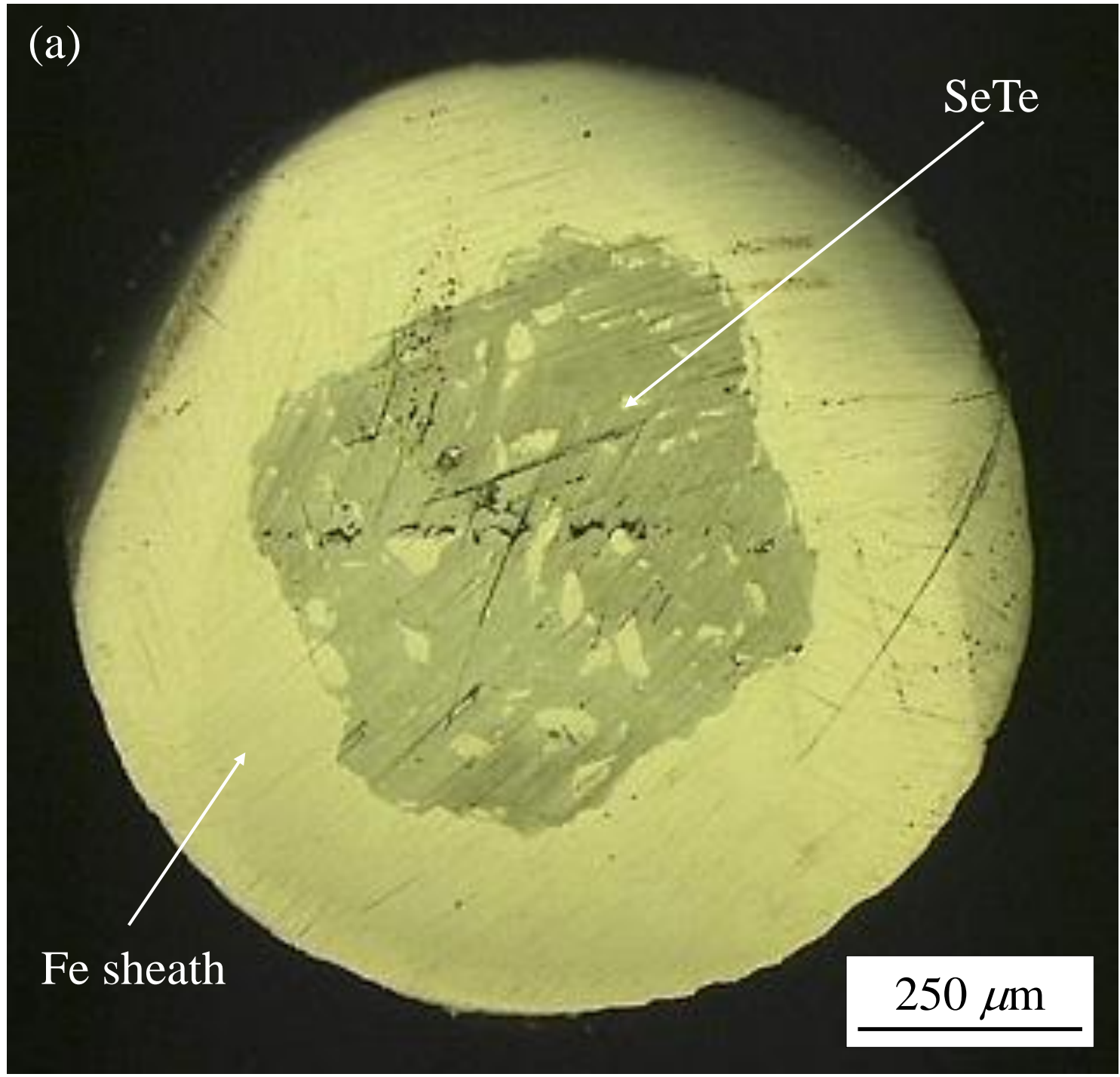

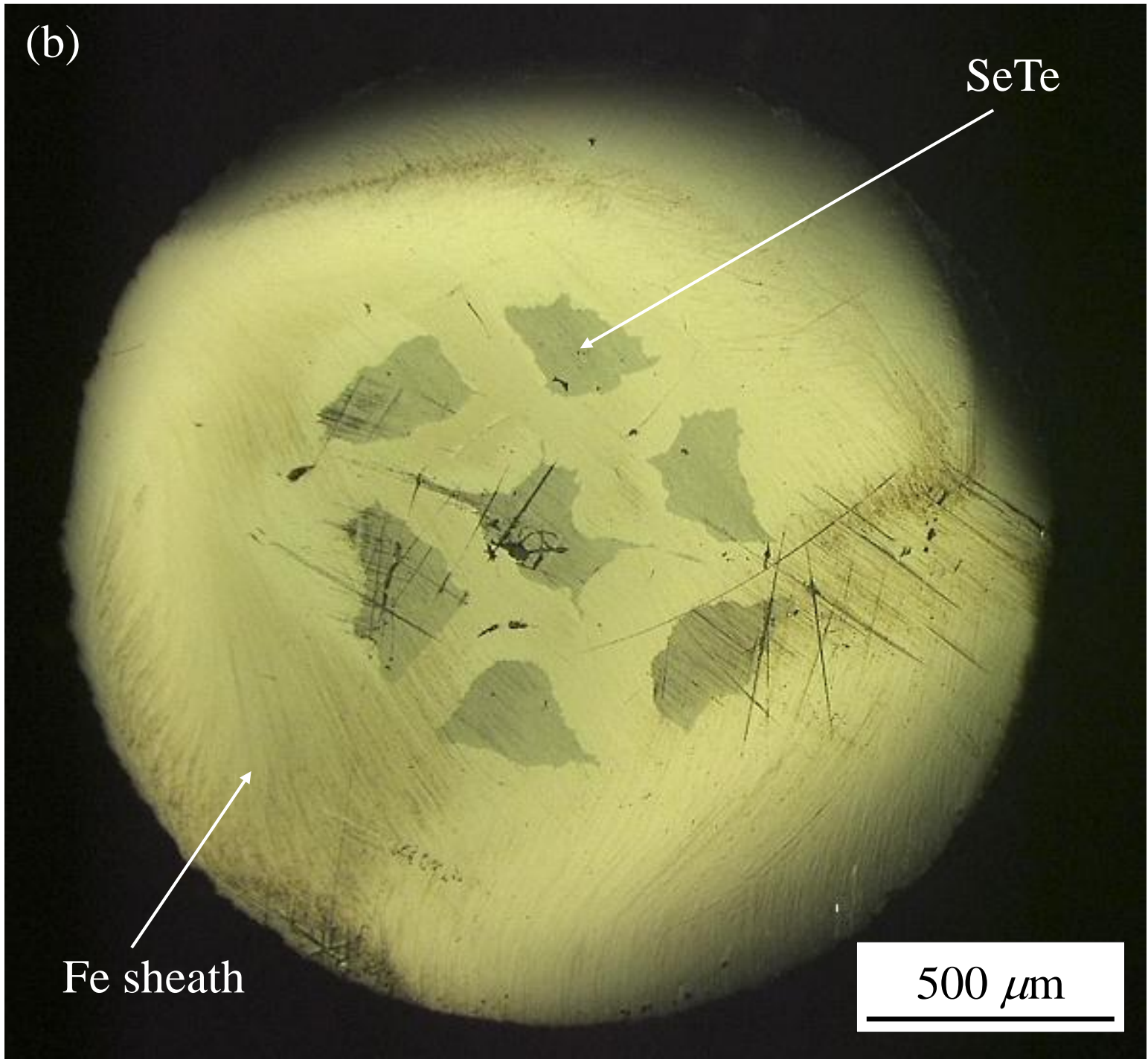

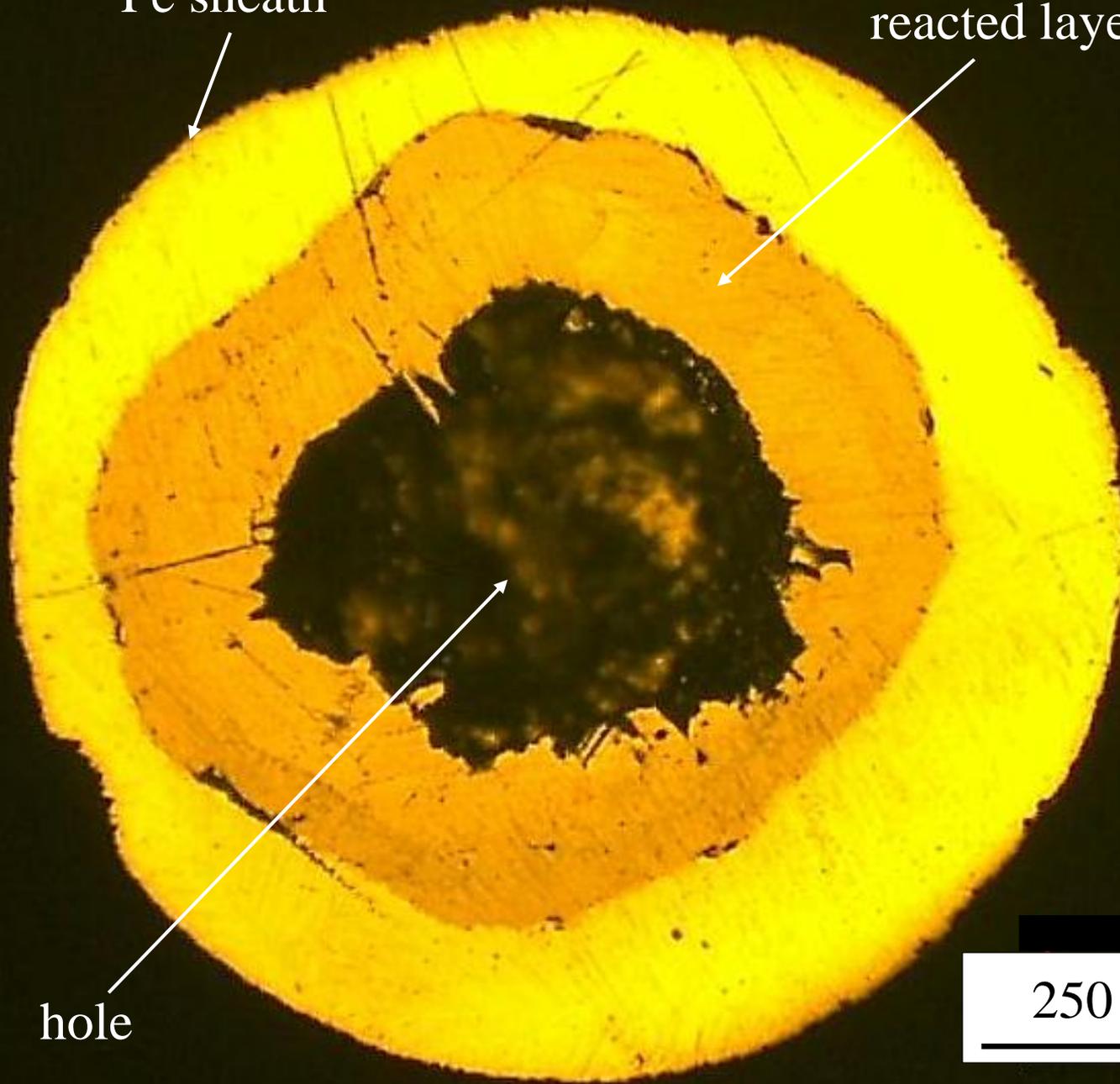

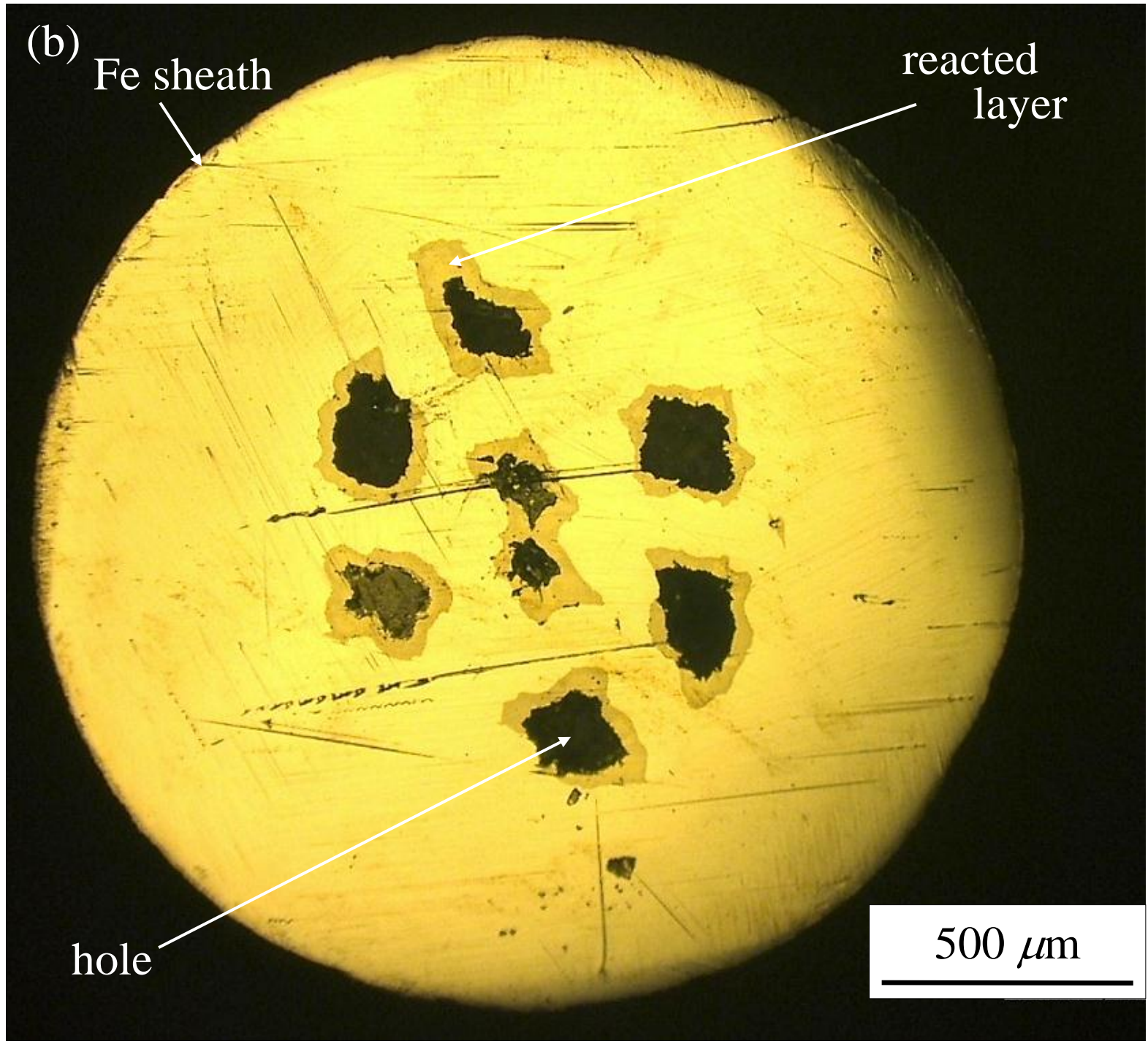

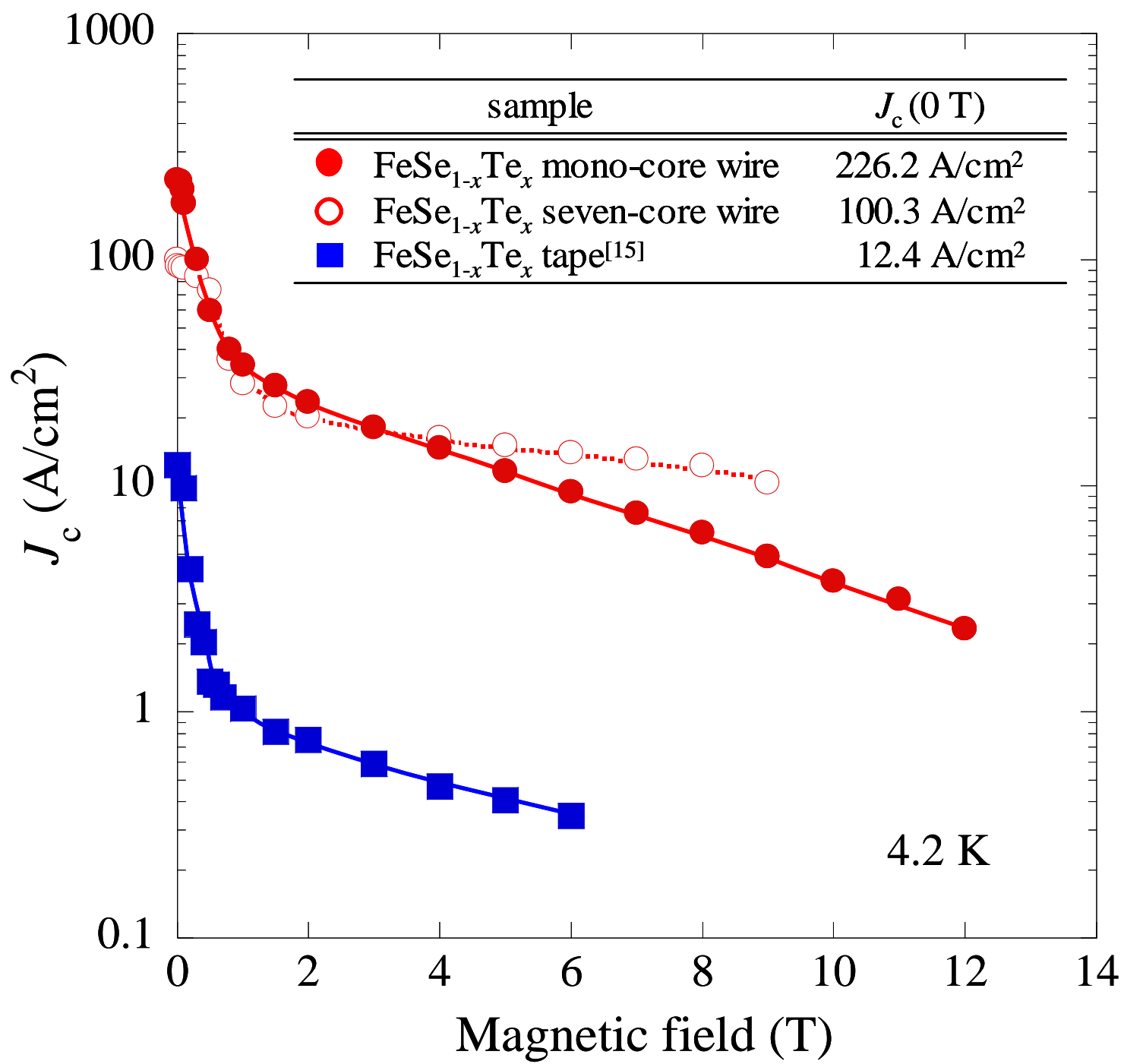

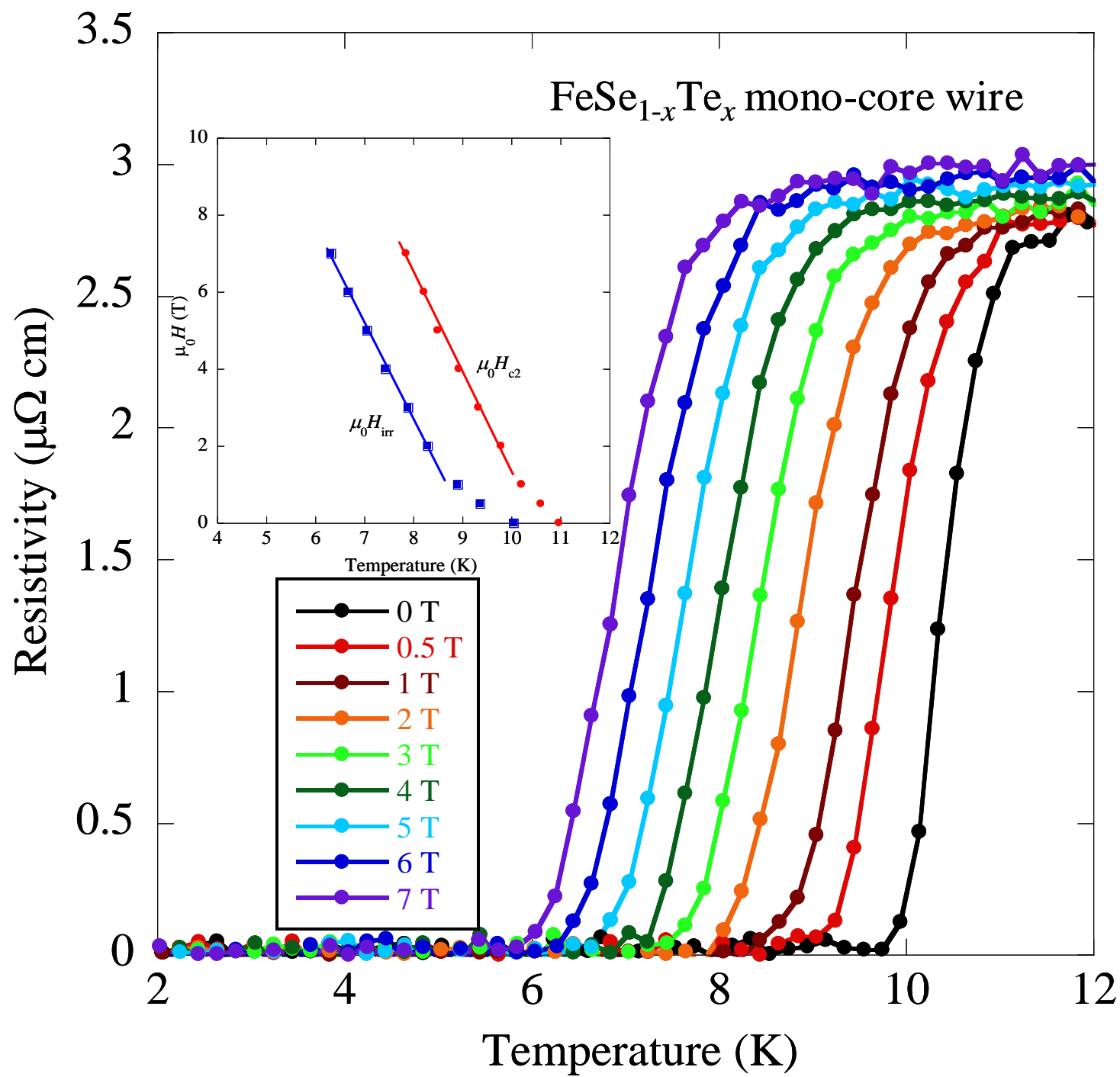

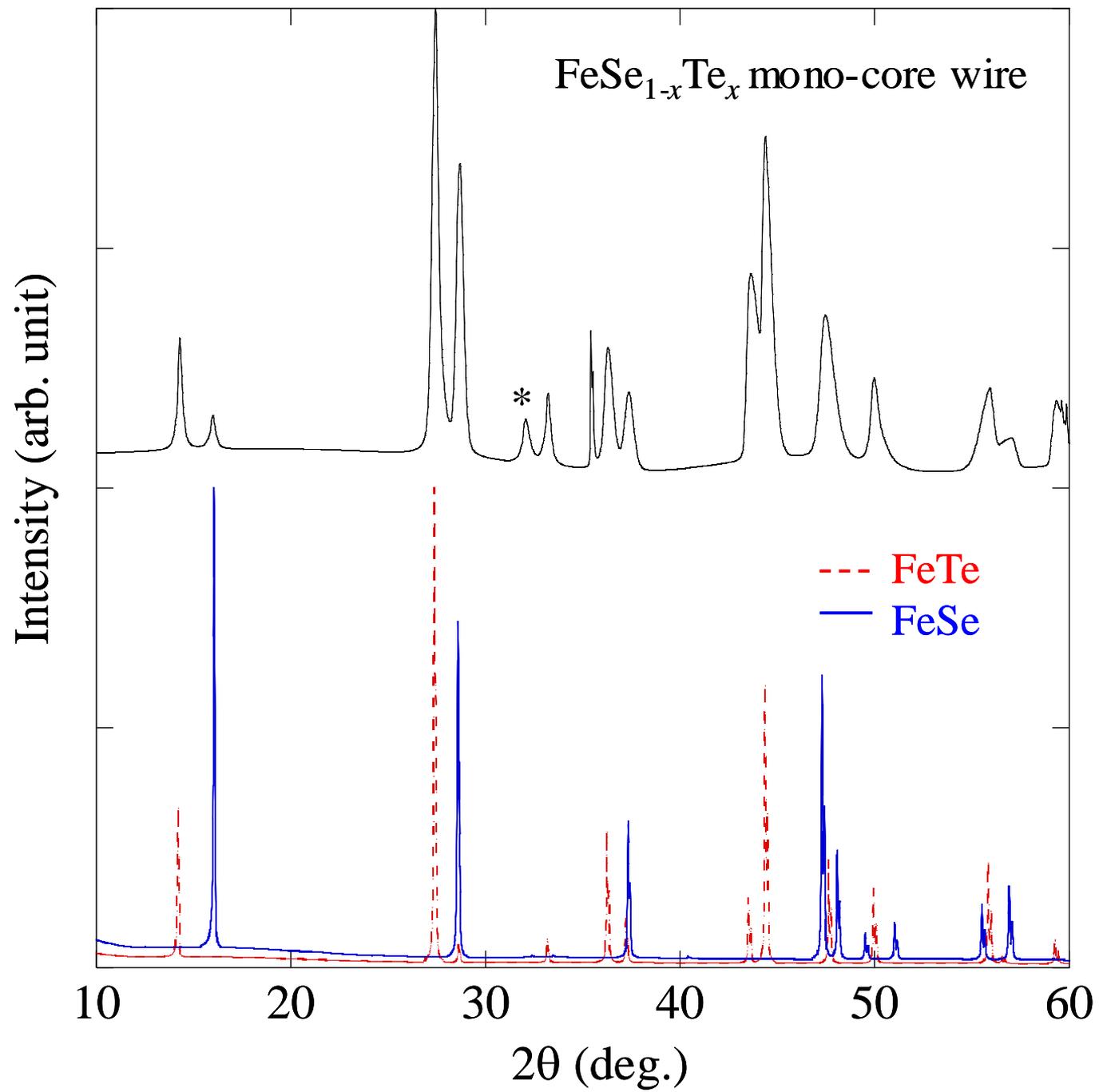

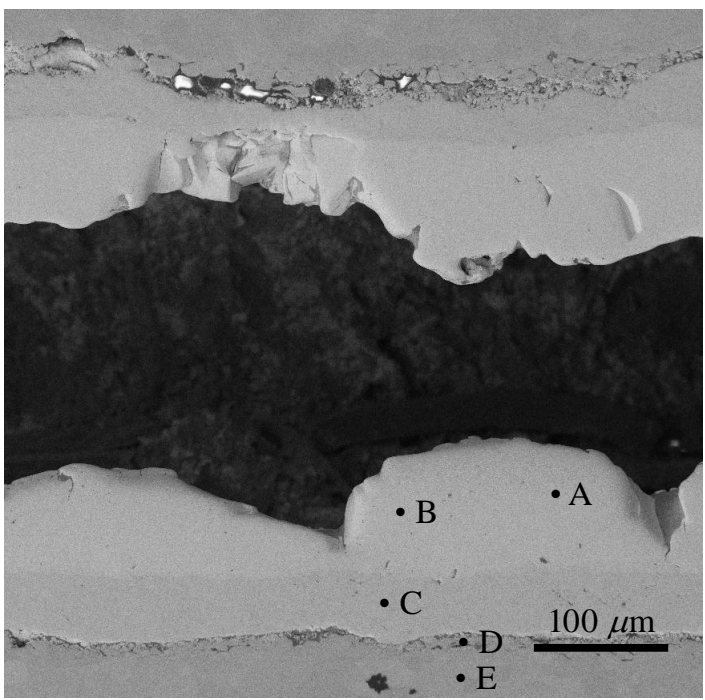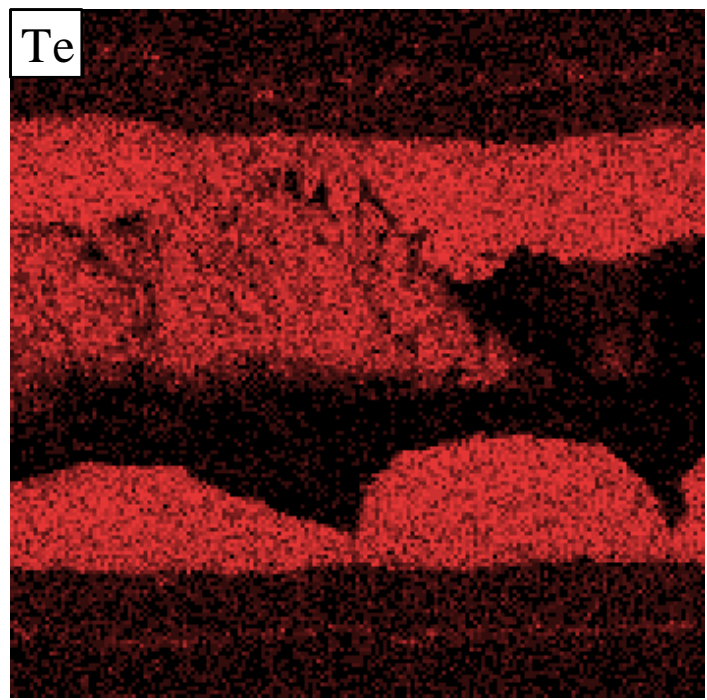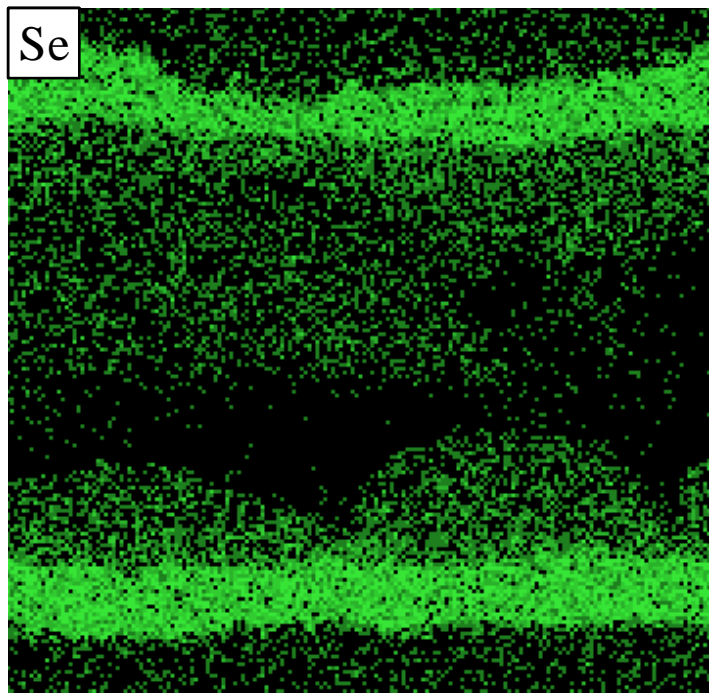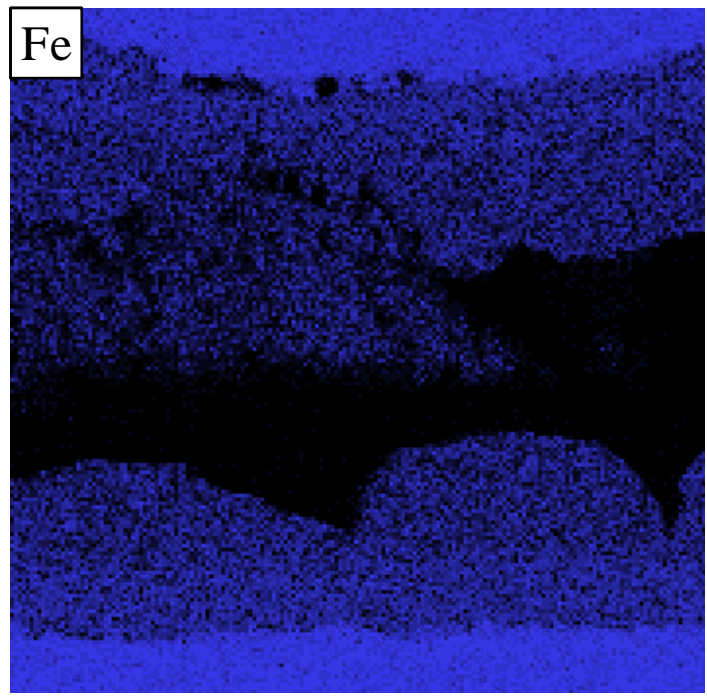